\documentclass[dvipsnames]{article} 

\usepackage{amsmath,amsfonts} 
\usepackage{array} 
\usepackage[caption=false,font=normalsize,labelfont=sf,textfont=sf]{subfig} 
\usepackage{textcomp} 
\usepackage{stfloats} 
\usepackage{url} 
\usepackage{verbatim} 
\usepackage{graphicx} 
\usepackage{booktabs} 
\usepackage{csquotes} 
\usepackage{longtable} 
\usepackage{balance} 
\usepackage{rotating} 
\usepackage[round]{natbib} 

\usepackage{tikz}
\usetikzlibrary{arrows.meta} 
\usetikzlibrary{shapes.geometric, positioning, fit, calc} 

\usepackage[
    pdfusetitle, 
    colorlinks=true,
    linkcolor=blue,
    filecolor=magenta,      
    urlcolor=cyan,
    citecolor=blue,
    bookmarks=true, 
    bookmarksopen=true
]{hyperref} 

\usepackage{orcidlink} 

\hypersetup{
    pdfkeywords={Large Language Models, Spatial Econometrics, Peer Review, Model Evaluation, Artificial Intelligence, GeoAI, Benchmarking, Scientific Validation}
}

\begin{document}

\title{Evaluating Large Language Model Capabilities in Assessing Spatial Econometrics Research}

\author{
    Giuseppe Arbia\thanks{Department of Statistical Sciences, Università Cattolica del Sacro Cuore, Rome, Italy. Email: giuseppe.arbia@unicatt.it. ORCID: \href{https://orcid.org/0000-0002-7926-1152}{\orcidlink{0000-0002-7926-1152}}}, 
    Luca Morandini\thanks{School of Computing and Information Systems, The University of Melbourne, Melbourne, Australia. Email: luca.morandini@unimelb.edu.au. ORCID: \href{https://orcid.org/0009-0009-3955-0396}{\orcidlink{0009-0009-3955-0396}}}, 
    Vincenzo Nardelli\thanks{Department of Statistical Sciences, Università Cattolica del Sacro Cuore, Rome, Italy. Email: vincenzo.nardelli@unicatt.it. ORCID: \href{https://orcid.org/0000-0002-7215-7934}{\orcidlink{0000-0002-7215-7934}}} 
}

\maketitle
\begin{abstract}
This paper investigates Large Language Models' (LLMs) ability to assess the economic soundness and theoretical consistency of empirical findings in spatial econometrics. We created original and deliberately altered "counterfactual" summaries from 28 published papers (2005-2024), which were evaluated by a diverse set of LLMs. The LLMs provided qualitative assessments and structured binary classifications on variable choice, coefficient plausibility, and publication suitability. The results indicate that while LLMs can expertly assess the coherence of variable choices (with top models like GPT-4o achieving an overall F1 score of 0.87), their performance varies significantly when evaluating deeper aspects such as coefficient plausibility and overall publication suitability. The results further revealed that the choice of LLM, the specific characteristics of the paper and the interaction between these two factors significantly influence the accuracy of the assessment, particularly for nuanced judgments. These findings highlight LLMs' current strengths in assisting with initial, more surface-level checks and their limitations in performing comprehensive, deep economic reasoning, suggesting a potential assistive role in peer review that still necessitates robust human oversight.
\end{abstract}


\section{Introduction}\label{sec:Introduction}
The rapid proliferation of Artificial Intelligence (AI), particularly Large Language Models (LLMs), is transforming numerous scientific domains. Their potential applications in research span from literature synthesis and hypothesis generation to data analysis assistance and code generation. A particularly compelling, yet challenging, area is the potential use of LLMs in the academic peer review process, a cornerstone of scientific validation that is often resource-intensive. Evaluating whether LLMs can reliably assist in assessing the quality and rigor of specialized research is crucial for understanding their future role in scholarly communication.

Spatial econometrics, a field focused on statistical methods tailored for data exhibiting spatial dependence and heterogeneity \citep{Anselin1988, arbia2014primer}, presents a unique test case. Models such as the spatial lag, spatial error, and spatial Durbin models incorporate complex spatial interactions, requiring nuanced interpretation of direct and indirect (spillover) effects \citep{LeSagePace2009}. Assessing the economic plausibility and theoretical consistency of the results derived from these models demand significant domain expertise. Can LLMs, trained on vast text corpora but potentially lacking deep causal reasoning, effectively evaluate the economic soundness of such specialized quantitative research?

This paper addresses this question by systematically evaluating an LLM's ability to perform a specific, critical component of peer review for spatial econometrics papers: assessing the economic interpretation and theoretical consistency of reported empirical results. We are not evaluating the LLM's capacity to judge novelty or the appropriateness of complex identification strategies, but rather its ability to act as an initial 'sense check' regarding the economic logic embedded within the model's findings.

To achieve this, we developed a rigorous methodology that involves the selection of published papers from prominent journals in the field. For each paper, we prepared standardized summaries that capture the objective of the core research, the variables, and the results. Crucially, we also generated a parallel set of ``counterfactual'' summaries where key results were deliberately modified to conflict with established economic theory. Then, both sets of summaries were evaluated by an LLM using a two-phase protocol: an initial qualitative assessment followed by a structured binary classification of key aspects of economic soundness and suitability.

By comparing the LLM's performance on summaries reflecting published (theoretically consistent) work versus those containing deliberate inconsistencies, we aim to quantify the LLM's ability to discriminate based on economic plausibility. This study provides empirical evidence on the current capabilities and limitations of LLMs in assisting the review process within a specialized quantitative field, providing insights for researchers, reviewers, and journal editors navigating the integration of AI into academic workflows.

The remainder of this paper is organized as follows: Section 2 reviews the background literature on spatial econometrics and the role of LLMs in research assessment; Section 3 details the methodology used for paper selection, summary creation, and the LLM evaluation protocol; Section 4 introduces the specific Large Language Models employed in the study; Section 5 presents the quantitative results of the LLM evaluations; Section 6 discusses the implications of these findings; and Section 7 concludes the paper, outlining limitations and suggesting future research directions.

\section{Background and Related Work}\label{sec:Background}
Spatial econometrics is a branch of econometrics that analyzes data where geography plays a crucial role. Unlike traditional approaches, it accounts for spatial dependence—nearby observations tend to be more similar, following Tobler’s First Law of Geography: “Everything is related to everything else, but near things are more related than distant things” \citep{tobler1970computer}. This recognition of spatial dependence is what differentiates spatial econometrics from classical econometric approaches and requires the use of specialized models and methods to avoid biased or inconsistent results \citep{arbia2014primer, anselin1995new}.

In recent years, Large Language Models (LLMs) have emerged as a powerful class of artificial intelligence (AI) systems capable of understanding and generating human language using deep learning techniques. Within scientific research, LLMs offer promising applications ranging from literature review and knowledge discovery to supporting data analysis and interpretation, hypothesis generation, and enabling natural language interfaces for scientific tools \citep{birhane2023science}.

Today, spatial econometrics is entering a new phase with advances in Artificial Intelligence, particularly Large Language Models (LLMs). LLMs are transforming scientific research by acting as intelligent assistants capable of processing, summarizing, and synthesizing vast scientific literature, accelerating knowledge discovery, and supporting complex analytical tasks. This technological shift raises critical questions and challenges for spatial econometrics. LLMs, while powerful, inherit biases from their training data \citep{pataranutaporn2025can} and require careful evaluation to ensure the reliability of their outputs \citep{zheng2023judging}. These developments call for a redefinition of the role of researchers and scientific associations, emphasizing the need for active monitoring, validation, and control of LLMs within the discipline.

The exponential growth in scientific publications has placed considerable strain on the traditional peer-review system, leading to well-documented challenges such as reviewer shortages, extended review timelines, and concerns over review consistency and quality \citep{hosseini2023fighting}. This situation is sometimes described as a ``peer review crisis'', prompting exploration into alternative or supplementary approaches to research assessment. In this context, Large Language Models (LLMs) have emerged as a technology with the potential to augment, or in some visions, partially automate aspects of the scientific evaluation process \citep{zhao2023survey}.

Initial research has explored the integration of various LLMs into the peer-review ecosystem, identifying potential applications ranging from initial manuscript screening and language checks to assisting reviewers with literature searches and generating preliminary feedback \citep{hosseini2023fighting, birhane2023science}. These models can generate reviews that appear plausible and sometimes overlap with comments made by human experts \citep{liang2024can}.

However, the capabilities of current LLMs in research assessment are subject to significant limitations. A consistent finding across studies is that while LLMs can mimic the style of academic reviews, they struggle with deep scientific understanding, particularly in evaluating methodological rigor, theoretical consistency, and the novelty of contributions \citep{shin2025automatically, naddaf2025ai}. They are prone to factual inaccuracies (hallucinations) \citep{lewis2020retrieval}, exhibit inconsistencies in their outputs \citep{zheng2023judging}, and can inherit and amplify biases present in their training data, such as those related to author affiliation or gender \citep{pataranutaporn2025can}. Furthermore, direct comparisons reveal low agreement between the substance of LLM-generated feedback and evaluations conducted by human domain experts \citep{shin2025automatically}.

While the general capabilities and limitations of LLMs in research assessment are becoming clearer, their performance within specialized, quantitative scientific domains requires specific investigation. Evaluating research in this area necessitates not only understanding statistical methodology but also assessing the economic soundness and theoretical consistency of the estimated spatial effects and relationships. The interpretation of parameters (e.g., spatial lags, spatial error terms, geographically weighted coefficients) requires domain-specific knowledge that goes beyond surface-level text analysis.

This paper addresses this gap by specifically evaluating the capability of an LLM to assess the theoretical consistency and economic meaningfulness of empirical findings within spatial econometrics. We hypothesize that the nuanced interpretation required in this field poses a significant challenge for current LLMs. To test this, we employ a novel evaluation strategy using paired original and deliberately falsified ``counterfactual'' summaries of published spatial econometrics papers. By removing author interpretations and focusing the LLM's task on the core results and their theoretical plausibility, we aim to provide a targeted assessment of its ability to perform a critical function of peer review in this specialized domain. This study seeks to understand whether LLMs can reliably distinguish between theoretically sound and unsound quantitative spatial research, thereby informing their potential utility and limitations as assistants in the peer-review process for fields like spatial econometrics.

\section{Methodology} \label{sec:Methodology}

This study employs a systematic, multi-stage methodology to evaluate the capability of a Large Language Model (LLM) to assess the economic soundness and theoretical consistency of applied spatial econometrics research. The process encompasses four core stages: targeted paper selection, standardized information extraction and summary creation, generation of counterfactual summaries, and a structured two-phase LLM evaluation protocol applied to both original and counterfactual summaries, as illustrated in Fig.~\ref{fig:workflow}.

\tikzset{
  block/.style = { 
    rectangle, 
    draw, 
    fill=blue!10,
    text width=8em, 
    text centered, 
    rounded corners, 
    minimum height=4em 
  },
  data/.style  = { 
    rectangle, 
    draw, 
    fill=green!10,
    text width=8em, 
    text centered, 
    rounded corners, 
    minimum height=4em 
  },
  process/.style= { 
    rectangle, 
    draw, 
    fill=orange!10,
    text width=8em, 
    text centered, 
    rounded corners, 
    minimum height=4em 
  },
  eval/.style   = { 
    rectangle, 
    draw, 
    fill=red!10,
    text width=8em, 
    text centered, 
    rounded corners, 
    minimum height=4em 
  },
  line/.style  = { 
    draw, 
    -{Stealth[length=3mm, width=2.5mm]} 
  }
}

\begin{figure*}[!t] 
    \centering
    \begin{tikzpicture}[node distance = 1cm and 3cm]
        \node [block] (select) {Paper Selection \& Sample Construction};

        \node [data, below=of select] (summary) {Standardized Summary Prep. (Original)};
        \node [process, below=of summary] (qual_orig) {Phase 1: Qualitative Eval. (Original)};
        \node [eval, below=of qual_orig] (struct_orig) {Phase 2: Structured Eval. (Original)};
        \node [block, below=of struct_orig] (analysis_orig) {Analysis (Original Results)};

        \node [data, right=of summary, node distance=7cm] (counter) {Counterfactual Summary Creation}; 
        \node [process, below=of counter] (qual_count) {Phase 1: Qualitative Eval. (Counterfactual)};
        \node [eval, below=of qual_count] (struct_count) {Phase 2: Structured Eval. (Counterfactual)};
        \node [block, below=of struct_count] (analysis_count) {Analysis (Counterfactual Results)};
        
        \node [block, below=of analysis_orig, yshift=-2cm, xshift=3.5cm ] (compare) {Comparative Analysis};

        \path [line] (select) -- (summary); 
        \path [line] (summary) -- node [midway, above, sloped, yshift=3mm, text width=5em] {Modify Results} (counter); 
        \path [line] (summary) -- (qual_orig); 
        \path [line] (counter) -- (qual_count); 
        \path [line] (qual_orig) -- (struct_orig); 
        \path [line] (qual_count) -- (struct_count); 
        \path [line] (struct_orig) -- (analysis_orig); 
        \path [line] (struct_count) -- (analysis_count); 
        
        \path [line] (analysis_orig) |- (compare); 
        \path [line] (analysis_count) |- (compare); 

        \node [draw, dashed, fit=(qual_orig) (qual_count) (struct_orig) (struct_count), inner sep=0.5cm, label={above:LLM Evaluation Protocol}] (llm_box) {}; 

    \end{tikzpicture}
    \caption{Methodological workflow for evaluating LLM capabilities in assessing spatial econometrics research. The process involves selecting papers, creating original and counterfactual summaries, and subjecting both to a two-phase LLM evaluation protocol before comparative analysis.}
    \label{fig:workflow} 
\end{figure*}
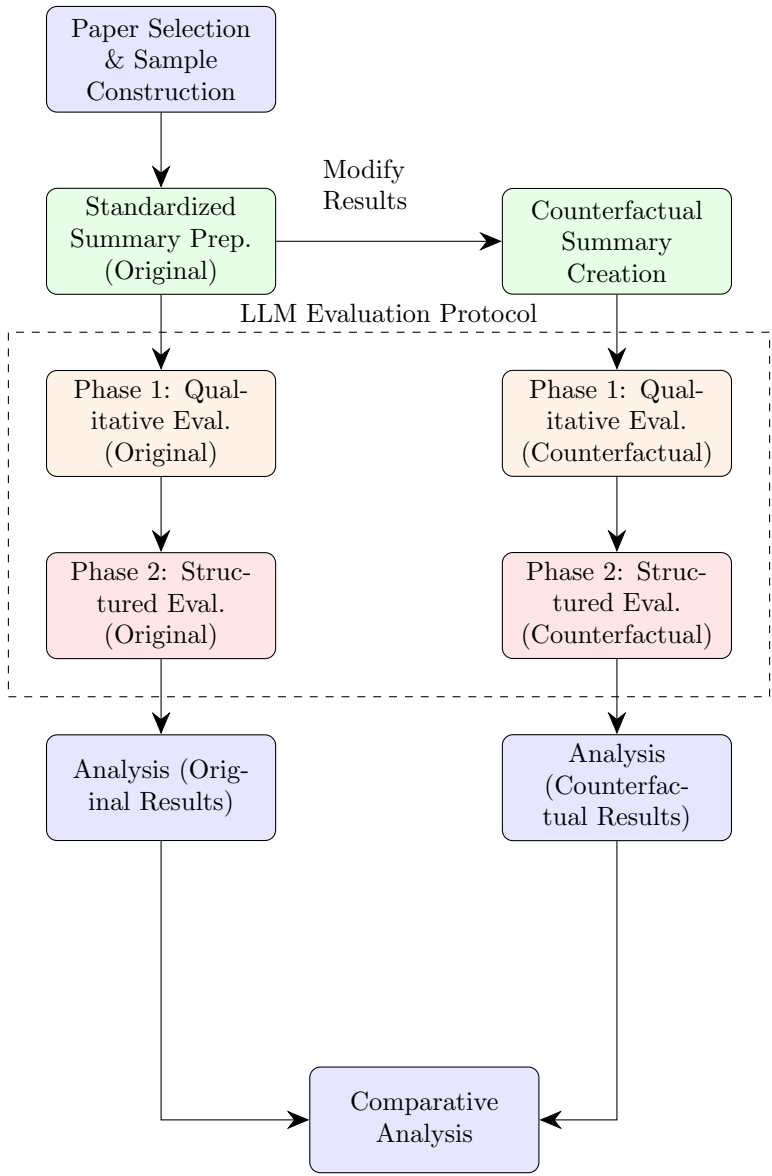

\subsection{Paper Selection and Sample Construction}

The first stage involved constructing a representative sample of relevant academic publications. A total of 28 applied spatial econometrics papers were selected from four leading peer-reviewed journals in the field: \textit{Journal of Applied Econometrics}, \textit{International Regional Science Review}, \textit{Spatial Economic Analysis}, and \textit{Papers in Regional Science}. The publication dates range from 2005 to 2024, capturing significant developments in the application of spatial techniques.

Selection criteria were carefully defined to ensure the relevance and suitability of the papers for the evaluation task. Specifically, papers had to feature empirical applications of established spatial econometric models, such as the spatial lag model (SAR) and spatial Durbin model (SDM). The focus was strictly on applied work where spatial models were used to derive substantive economic insights from data, explicitly linking findings to economic theory. A critical requirement was the clear presentation of quantitative results; each selected paper needed to contain at least one detailed table, either in the main text or an appendix, reporting estimated model coefficients alongside robust indicators of statistical significance (e.g., p-values, standard errors, or significance stars). Theoretical contributions and simulation studies were excluded. The final selection involved manual curation to confirm adherence to all criteria and to identify the primary results table representing the core economic findings intended by the authors.

\subsection{Preparation of Standardized Summaries for LLM Input}

To ensure a controlled and unbiased evaluation by the LLM, standardized summaries were meticulously prepared for each selected paper based on the published findings. This step aimed to provide the LLM with sufficient context and the core results needed for an economic assessment, while deliberately omitting potentially confounding information. Each summary was structured into three distinct components:
\begin{enumerate}
    \item \textbf{Research Aim:} A concise statement articulating the central research question or objective pursued in the paper. This provides the necessary context regarding the purpose of the econometric model.
    \item \textbf{Variable Descriptions:} Precise definitions for the dependent variable and all explanatory variables included in the selected results table. Information regarding units of measurement or the specific nature of variables was included where pertinent.
    \item \textbf{Model Results Table:} A faithful transcription or reformatting of the key table from the original paper, presenting the variable names, their estimated coefficients, and the associated statistical significance indicators as published.
\end{enumerate}
Crucially, these summaries intentionally excluded the paper's identifying metadata (title, authors, journal), any interpretations or discussion of the results provided by the original authors, and all citations or references. This design isolates the evaluation task, compelling the LLM to base its assessment solely on the stated aim, the variables employed, and the quantitative results presented in the table, thereby simulating a focused review of the model's intrinsic economic logic.

\subsection{Creation of Counterfactual Summaries}

Following the creation of summaries based on the original published papers, a parallel set of ``counterfactual'' summaries was generated. For each paper, the results table within its standardized summary underwent deliberate modification. This involved manually altering the sign of key coefficients or adjusting the statistical significance of theoretically pivotal variables. These targeted edits were designed to introduce specific deviations from established economic theory, effectively creating versions of the results that exhibit theoretical inconsistencies or implausibility, while preserving the original research aim and variable descriptions. The objective of generating this counterfactual dataset was to establish a controlled basis for assessing the LLM's capacity to differentiate between model results consistent with economic theory (represented by the published findings) and those deliberately constructed to be theoretically inconsistent. This comparative approach is fundamental to evaluating the LLM's sensitivity to the economic plausibility of empirical results.

\subsection{LLM Evaluation Protocol}

The core of the methodology lies in the two-phase protocol designed to elicit and structure the LLM's evaluation, applied independently to both the original and the counterfactual summaries for each paper.

\textbf{Phase 1: Qualitative Economic Assessment.} In the initial phase, the LLM was prompted (using the same instructions for both original and counterfactual summaries) to conduct a qualitative review of the economic aspects of the model based on the provided summary. The LLM was instructed to adopt the persona of an expert in applied econometrics and spatial economics. Its core task involved performing a detailed economic assessment by first interpreting the relevant parameters, explaining the economic meaning of key coefficients, particularly those central to the research aim or exhibiting statistical significance, including spatial parameters like autoregressive coefficients ($\rho$) or spillover effects where applicable. Subsequently, the LLM evaluated the economic meaningfulness of the overall model specification, assessing whether the choice of variables represented a plausible economic relationship given the stated research aim. Furthermore, it assessed the theoretical consistency of the findings, evaluating if the signs and statistical significance of coefficients aligned with established economic theory or intuition, checking, for instance, if variables expected to have positive impacts showed significant positive coefficients. Finally, the LLM was tasked with identifying any inconsistencies or surprises, highlighting findings that contradicted theory, deviated from common empirical results, appeared counterintuitive, or seemed plausible but lacked strong theoretical backing in the provided context, thus requiring further author justification. (The full prompt is detailed in the Appendix.) This initial qualitative step was designed not only to elicit detailed feedback but also to encourage the LLM to engage in a more thorough reasoning process regarding the economic nuances of the model before proceeding to the structured summary judgment in the subsequent phase. This phase generates a narrative output detailing the LLM's reasoning for each summary type (original and counterfactual).

\textbf{Phase 2: Structured Evaluation Output.} Following the detailed qualitative assessment in Phase 1, a second phase was introduced to systematically capture the LLM's summary judgments in a quantifiable manner for each evaluated summary (original and counterfactual). This phase required the LLM to condense its evaluation into clear, binary classifications (1/0) across three distinct dimensions of economic soundness (The specific prompt is detailed in the Appendix). The rationale for this step was twofold: firstly, to obtain specific judgments on critical aspects of the economic evaluation, and secondly, to generate data suitable for quantitative analysis, such as the construction of an accuracy matrix comparing LLM performance against the known nature (published vs. counterfactual) of the summaries.

Specifically, the LLM provided classifications assessing the foundational logic of the model specification by evaluating the \textbf{economic sense of the variable choice}, determining if the selection appeared logical and justified within the economic context of the research aim. It also focused on the plausibility of the core empirical results by judging the \textbf{economic sense of the significant coefficients}, assessing whether the statistically significant findings (signs and magnitudes) aligned with established economic theory or reasonable expectations. Finally, the LLM offered an overall assessment by classifying the \textbf{suitability for publication based on scientific rigor and journal fit}, judging if the analysis, based purely on the provided rationale and results, demonstrated a level of scientific rigor and alignment with the implicit standards and scope of the relevant journals.

This structured approach allows for a clear and simplified summary of the LLM's assessment on these key criteria for both sets of summaries. However, this simplification also presents potential drawbacks: reducing complex judgments to binary outputs inevitably leads to a loss of nuance present in the qualitative evaluation, and the assessment of ``suitability for publication'' is inherently limited as it is based solely on the provided summary, lacking the full context of the original paper. Nonetheless, this phase yields quantifiable data points crucial for systematically analyzing and comparing the LLM's performance across the original and counterfactual datasets for the entire sample of papers.

\section{Large Language Models Used} \label{sec:LLMsUsed}

This section summarizes the Large Language Models (LLMs) selected to evaluate their ability to assess economic soundness and theoretical consistency in spatial econometrics research. The chosen models represent a diverse range from developers including OpenAI, xAI, Meta, Google, DeepSeek, and Anthropic, encompassing various scales, architectures, licensing models (proprietary vs. open/community), and specialized capabilities like advanced reasoning and large context windows. Understanding these differences is key to interpreting the study's findings on LLM performance in academic review tasks.

The models include OpenAI's generalist GPT-4o \citep{openai:gpt4o_img} and the reasoning-focused o1 \citep{openai:o1pro_platform, openai:o1_syscard}; xAI's Grok-2 \citep{xai:docs_overview, xai:docs_models}; Meta's open-community Llama-3.3-70B \citep{llama:llama3.3_docs}; Google's Gemini-2.5-Flash Preview \citep{google:gemini2.5_dev}; DeepSeek's Mixture-of-Experts (MoE) models (DeepSeek-R1 reasoner \citep{github:deepseekr1}, DeepSeek-Chat-V3 generalist \citep{github:deepseekv3, deepseek:v3_news}) and a distilled 8B model (DeepSeek-R1-Distill-8B) \citep{github:deepseekr1}; and Anthropic's Claude 3.7 Sonnet featuring hybrid reasoning capabilities \citep{anthropic:claude3.7_news, anthropic:claude3.7_syscard}. Key variations include context window sizes ranging from 128k upwards, diverse licensing from proprietary APIs to open weights, and specific optimizations for reasoning versus general-purpose use.

\subsection*{LLM Architectures}

The models employed utilize different underlying architectures, influencing their capabilities and efficiency. Here are brief descriptions, ordered by increasing complexity:

\begin{itemize}
    \item \textbf{Dense Transformer Architecture:} This is the foundational architecture for many LLMs. In each layer, every input token interacts with every other token (self-attention), and typically, all model parameters are used during computation for each token. While powerful, computational cost scales significantly with input length and model size. Meta's Llama models are examples of this architecture \citep{llama:llama3.3_docs}. The DeepSeek distilled model also uses this architecture as its base \citep{github:deepseekr1}.
    \item \textbf{Mixture-of-Experts (MoE) Architecture:} This architecture aims for greater efficiency at scale. It consists of numerous specialized "expert" sub-networks within its layers. For each input token, a routing mechanism activates only a small subset of these experts (e.g., 1 or 2 out of many). This allows the model to have a very large total parameter count, potentially capturing vast knowledge, while keeping the computational cost per token lower than a dense model of equivalent total size. DeepSeek's R1 and V3 models use this approach \citep{github:deepseekr1, github:deepseekv3}.
    \item \textbf{Unified Architecture:} This term often describes models designed to natively handle multiple data types (e.g., text, images, audio) within a single, integrated network, rather than relying on separate specialized components linked together. OpenAI's GPT-4o is described as having a unified architecture for its multimodal capabilities \citep{openai:gpt4o_img}.
    \item \textbf{Hybrid Reasoning Architecture:} This represents an advanced design or operational mode focused on enhancing complex problem-solving. Models with this feature, like Anthropic's Claude 3.7 Sonnet, can dynamically allocate more computational resources and employ more explicit, potentially transparent, step-by-step reasoning processes when tackling difficult tasks, often offering a trade-off between speed/cost and analytical depth/rigor \citep{anthropic:claude3.7_news, anthropic:claude3.7_syscard}. Google's description of some Gemini models as "thinking models" may point towards similar principles \citep{google:gemini_blog}.
\end{itemize}

\subsection*{Comparative Table}
The following table summarizes key characteristics of the LLMs, focusing on architecture where specified.

\begin{sidewaystable}
    \centering 
    \begin{longtable}{@{} >{\raggedright\arraybackslash}p{4.0cm} p{2.5cm} >{\raggedright\arraybackslash}p{2.5cm} >{\raggedright\arraybackslash}p{2.0cm} >{\raggedright\arraybackslash}p{2.5cm} >{\raggedright\arraybackslash}p{3.0cm} @{}}
    \caption{Key Characteristics of Large Language Models Employed in the Study}\label{tab:llm_characteristics}\\
    \toprule
    \multicolumn{1}{c}{\bfseries Model Name} &
    \multicolumn{1}{c}{\bfseries Developer} &
    \multicolumn{1}{c}{\bfseries Parameter} &
    \multicolumn{1}{c}{\bfseries Context Window (Tokens)} &
    \multicolumn{1}{c}{\bfseries License Type} &
    \multicolumn{1}{c}{\bfseries Architecture} \\
    \midrule
    \endfirsthead
    \caption{Key Characteristics of Large Language Models Employed in the Study (Continued)}\\
    \toprule
    \multicolumn{1}{c}{\bfseries Model Name} &
    \multicolumn{1}{c}{\bfseries Developer} &
    \multicolumn{1}{c}{\bfseries Parameter} &
    \multicolumn{1}{c}{\bfseries Context Window (Tokens)} &
    \multicolumn{1}{c}{\bfseries License Type} &
    \multicolumn{1}{c}{\bfseries Architecture} \\
    \midrule
    \endhead
    \midrule
    \multicolumn{6}{r@{}}{\textit{Continued on next page}} \\
    \endfoot
    \bottomrule
    \endlastfoot
    GPT-4o\textsuperscript{a} & OpenAI & NS & 128k & Proprietary & Unified \citep{openai:gpt4o_img} \\
    \addlinespace
    o1\textsuperscript{b} & OpenAI & NS & 200k & Proprietary & NS \\
    \addlinespace
    Grok-2\textsuperscript{c} & xAI & NS & 128k & Proprietary & NS \\
    \addlinespace
    Llama-3.3-70B\textsuperscript{d} & Meta & 70B & 128k & Llama 3.3 Community Lic. & Dense Transformer \citep{llama:llama3.3_docs} \\
    \addlinespace
    Gemini-2.5-Flash Preview\textsuperscript{e} & Google & NS & NS & Proprietary & NS \\ 
    \addlinespace
    DeepSeek-R1\textsuperscript{f} & DeepSeek AI & 671B & 128k & NS & MoE \citep{github:deepseekr1} \\
    \addlinespace
    DeepSeek-R1-Distill-8B\textsuperscript{g} & DeepSeek AI & 8B & 128k & Llama 3.1 Community Lic. & Dense Transformer \citep{github:deepseekr1} \\
    \addlinespace
    DeepSeek-Chat-V3\textsuperscript{h} & DeepSeek AI & 671B & 128k & MIT License & MoE \citep{github:deepseekv3, deepseek:v3_news} \\
    \addlinespace
    Claude-3.7-Sonnet\textsuperscript{i} & Anthropic & NS & 200k & Proprietary & Hybrid Reasoning \citep{anthropic:claude3.7_news, anthropic:claude3.7_syscard} \\
    \end{longtable}

    \footnotesize 
    \textbf{Table Notes:} NS = Not Specified. \\
    \textsuperscript{a}Refers to `gpt-4o`. \textsuperscript{b}Refers to `o1`. \textsuperscript{c}Refers to `x-ai/grok-2-1212`. \textsuperscript{d}Refers to `meta-llama/Llama-3.3-70b-instruct`. \textsuperscript{e}Refers to `google/gemini-2.5-flash-preview:thinking`. \textsuperscript{f}Refers to `deepseek/deepseek-r1`. \textsuperscript{g}Refers to `deepseek/deepseek-r1-distill-llama-8b`. \textsuperscript{h}Refers to `deepseek/deepseek-chat-v3-0324`. \textsuperscript{i}Refers to `anthropic/claude-3.7-sonnet`.
    \normalsize 
\end{sidewaystable}

\clearpage

\section{Results}\label{sec:Results}

This section presents the quantitative findings from the structured evaluation (Phase 2) of the Large Language Models' (LLMs) assessment of original and counterfactual summaries for the 28 spatial econometrics papers. Performance is analyzed using Precision, Recall, and F1 Score across three criteria: economic sense of variable choice, economic sense of significant coefficients, and suitability for publication. We also report overall performance metrics, the distribution of processing failures, and ANOVA results to assess the significance of different factors influencing LLM performance. The specific LLMs evaluated are detailed in Table \ref{tab:llm_characteristics}.

\subsection{Distribution of Failures}
The reliability of the LLMs in producing the structured JSON output required in Phase 2 of the evaluation was assessed. Table \ref{tab:failure_rate} shows the percentage of instances where models failed to return the requested JSON output or returned a malformed/incomplete JSON object. Most models exhibited high reliability, with Claude-3.7-Sonnet, DeepSeek-R1, Gemini-2.5-Flash Preview, GPT-4o, o1, and Grok-2 achieving a less than 1.00\% failure rate. DeepSeek-Chat-V3 had a minimal failure rate of 1.21\%, whilst the models with a smaller number of parameters had a higher failure rate, with Llama-3.3-70B at 2.93\%, and DeepSeek-R1-Distill-Llama-8B at 13.62\%.

These failures typically involved deviations from the strict JSON format requested or omission of some fields. Such issues could stem from several factors: the inherent stochasticity of LLM outputs, the complexity of translating a nuanced qualitative assessment (Phase 1) into a constrained binary format (Phase 2), or the specific model's training and fine-tuning for structured data generation and adherence to complex instructions. Models not extensively optimized for precise JSON output might be more prone to these errors, especially when the input prompts are long or the cognitive load of the preceding task is high. In our pipeline, these failures were minimized by robust error handling, parsing checks, and re-querying mechanisms to improve the quality of results.

\begin{table}[!htbp]
\centering
\caption{Distribution of failures by LLM}
\label{tab:failure_rate}
\begin{tabular}[t]{@{}ll@{}}
\toprule
\multicolumn{1}{c}{\textbf{LLM}} & \multicolumn{1}{c}{\textbf{Failure rate}}\\
\midrule
Claude-3.7-Sonnet & 0.17\%\\
DeepSeek-Chat-V3 & 1.21\%\\
DeepSeek-R1 & 0.52\%\\
DeepSeek-R1-Distill-8B & 13.62\%\\
Gemini-2.5-Flash Preview & 0.34\%\\
GPT-4o & 0.52\%\\
Grok-2 & 0.17\%\\
Llama-3.3-70B & 2.93\%\\
o1 & 0.34\%\\
\bottomrule
\end{tabular}
\end{table}

\subsection{Performance on Variable Selection}
\begin{figure}[!htbp]
    \centering
     \includegraphics[width=\columnwidth]{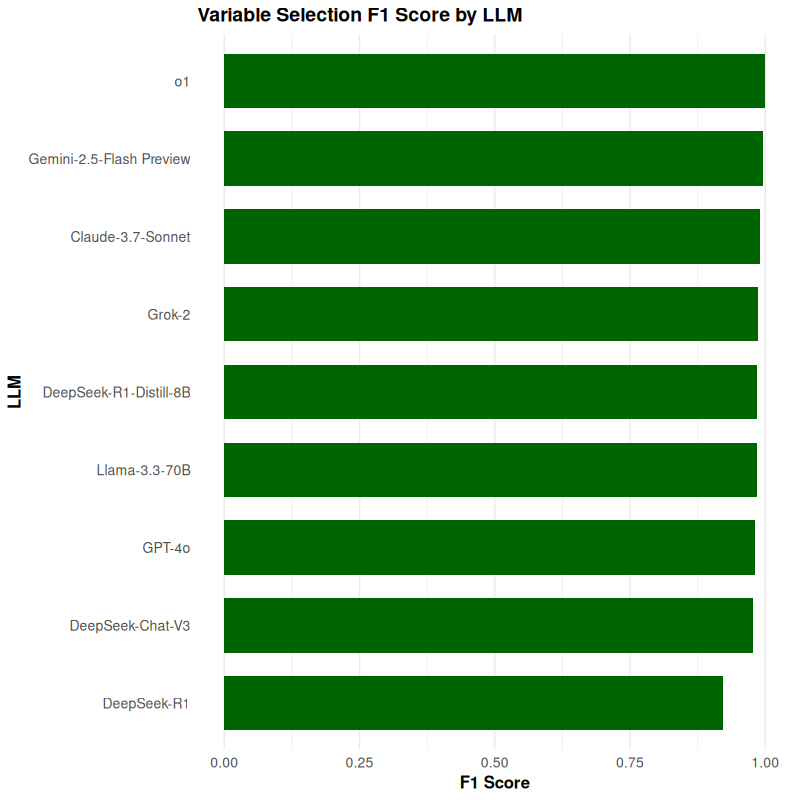} 
    \caption{F1 Scores for Variable Selection by LLM.}
    \label{fig:var_f1_score}
\end{figure}

The ability of LLMs to assess the economic sense of variable choices was evaluated. Detailed metrics are presented in Table \ref{tab:var_selection_metrics_detailed}.
Precision for this task was perfect (1.00) for all LLMs. Recall scores were also very high, ranging from 0.85 (DeepSeek-R1) to near 1.00 (o1, Gemini-2.5-Flash Preview).
The F1 scores, also listed in Table \ref{tab:var_selection_metrics_detailed} and visualized in Figure \ref{fig:var_f1_score}, were consequently excellent.

\begin{table}[!htbp]
\centering
\caption{Precision, Recall, and F1 Score for Variable Selection}
\label{tab:var_selection_metrics_detailed}
\begin{tabular}[t]{@{}lrrr@{}}
\toprule
\multicolumn{1}{c}{\textbf{LLM}} & \multicolumn{1}{c}{\textbf{Precision}} & \multicolumn{1}{c}{\textbf{Recall}} & \multicolumn{1}{c}{\textbf{F1 Score}}\\
\midrule
Claude-3.7-Sonnet & 1 & 0.98 & 0.99\\
DeepSeek-Chat-V3 & 1 & 0.96 & 0.98\\
DeepSeek-R1 & 1 & 0.85 & 0.92\\
DeepSeek-R1-Distill-8B & 1 & 0.97 & 0.98\\
GPT-4o & 1 & 0.96 & 0.98\\
Gemini-2.5-Flash Preview & 1 & 0.99 & 1.00\\
Grok-2 & 1 & 0.97 & 0.99\\
Llama-3.3-70B & 1 & 0.97 & 0.98\\
o1 & 1 & 1.00 & 1.00\\
\bottomrule
\end{tabular}
\end{table}

The ANOVA results for variable selection success are presented in Table \ref{tab:anova_var_selection}. All factors—the specific LLM used (`llm`), the paper summary being evaluated (`paper`), and the interaction between them (`llm:paper`)—were found to be highly statistically significant (p < 0.001). This indicates that: (1) there are significant differences in performance among the LLMs for this task; (2) the characteristics of the paper summary itself significantly affect how well LLMs can assess variable choice; and (3) crucially, the performance of a particular LLM varies depending on the specific paper summary it is evaluating, and this variation pattern differs across LLMs. Despite the overall high performance, these significant effects suggest nuances in how different LLMs approach even this relatively straightforward task depending on the input.

\begin{table}[!htbp]
\centering
\caption{Variable Selection ANOVA Results}
\label{tab:anova_var_selection}
\begin{tabular}[t]{@{}lrrrrl@{}}
\toprule
Source & df & Sum of Squares & Mean Square & F-value & p-value\\
\midrule
llm & 8 & 15.583 & 1.948 & 44.245 & 5.98e-69***\\
model & 57 & 15.429 & 0.271 & 6.148 & 6.79e-42***\\
llm:model & 456 & 53.746 & 0.118 & 2.677 & 8.68e-60***\\
Residuals & 4698 & 206.832 & 0.044 & NA & NA\\
\bottomrule
\multicolumn{6}{l}{\rule{0pt}{1em}\textit{Significance levels: } *** p $<$ 0.001}\\
\end{tabular}
\end{table}

\subsection{Performance on Coefficient Plausibility}
\begin{figure}[!htbp]
    \centering
    \includegraphics[width=\columnwidth]{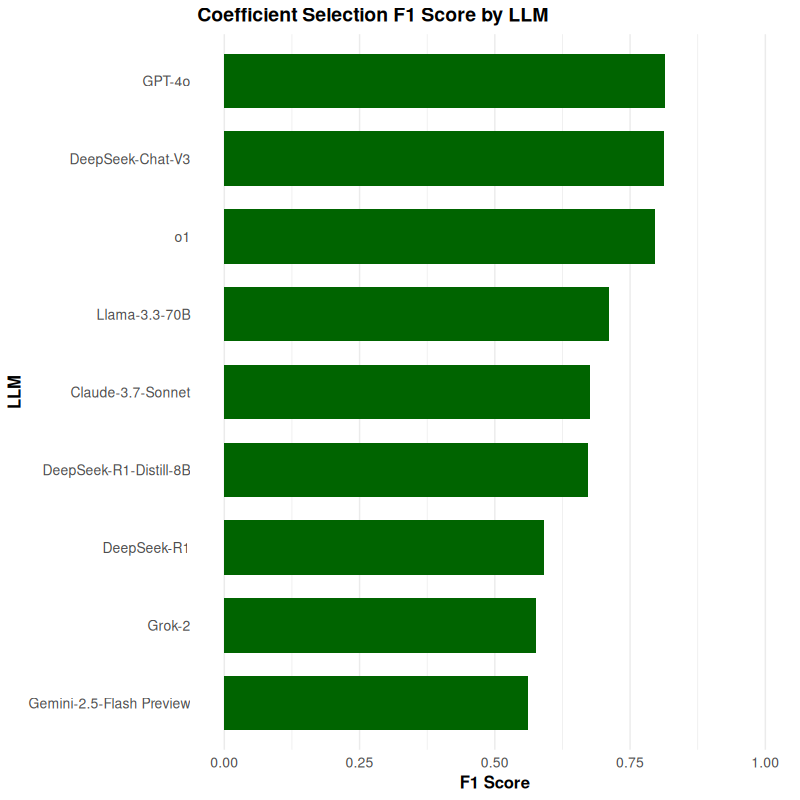}
    \caption{F1 Scores for Coefficient Selection by LLM.}
    \label{fig:coeff_f1_score}
\end{figure}

Assessing the economic sense of significant coefficients proved more challenging, with metrics in Table \ref{tab:coeff_selection_metrics_detailed}.
Precision varied, with DeepSeek-R1 (0.84) and Gemini-2.5-Flash Preview (0.85) showing higher precision. Recall also varied, with o1 (0.97) and DeepSeek-R1-Distill-8B (0.95) performing well.
The F1 scores (Table \ref{tab:coeff_selection_metrics_detailed} and Figure \ref{fig:coeff_f1_score}) show GPT-4o (0.81), DeepSeek-Chat-V3 (0.81), and o1 (0.80) as top performers.

\begin{table}[!htbp]
\centering
\caption{Precision, Recall, and F1 Score for Coefficient Selection}
\label{tab:coeff_selection_metrics_detailed}
\begin{tabular}[t]{@{}lrrr@{}}
\toprule
\multicolumn{1}{c}{\textbf{LLM}} & \multicolumn{1}{c}{\textbf{Precision}} & \multicolumn{1}{c}{\textbf{Recall}} & \multicolumn{1}{c}{\textbf{F1 Score}}\\
\midrule
Claude-3.7-Sonnet & 0.79 & 0.59 & 0.68\\
DeepSeek-Chat-V3 & 0.76 & 0.87 & 0.81\\
DeepSeek-R1 & 0.84 & 0.45 & 0.59\\
DeepSeek-R1-Distill-8B & 0.52 & 0.95 & 0.67\\
GPT-4o & 0.75 & 0.89 & 0.81\\
Gemini-2.5-Flash Preview & 0.85 & 0.42 & 0.56\\
Grok-2 & 0.78 & 0.46 & 0.58\\
Llama-3.3-70B & 0.78 & 0.65 & 0.71\\
o1 & 0.67 & 0.97 & 0.80\\
\bottomrule
\end{tabular}
\end{table}

The ANOVA results for coefficient selection success (Table \ref{tab:anova_coeff_selection}) again show that `llm`, `paper`, and their interaction `llm:paper` are all highly significant (p $<$ 0.001). The F-values are notably larger for this criterion compared to variable selection, particularly for the `llm` and `paper` main effects. This reinforces that LLM choice matters significantly, that some paper summaries are inherently harder to evaluate correctly in terms of coefficient plausibility, and that the interaction effect is strong: the relative performance of LLMs changes depending on the specific paper summary. This task, requiring deeper economic reasoning, elicits more pronounced differences and context dependencies in LLM performance.

\begin{table}[!htbp]
\centering
\caption{Coefficient Selection ANOVA Results}
\label{tab:anova_coeff_selection}
\begin{tabular}[t]{@{}lrrrrl@{}}
\toprule
Source & df & Sum of Squares & Mean Square & F-value & p-value\\
\midrule
llm & 8 & 40.877 & 5.110 & 56.428 & 3.74e-88***\\
model & 57 & 215.856 & 3.787 & 41.821 & 0.00e+00***\\
llm:model & 456 & 424.322 & 0.931 & 10.276 & 0.00e+00***\\
Residuals & 4698 & 425.415 & 0.091 & NA & NA\\
\bottomrule
\multicolumn{6}{l}{\rule{0pt}{1em}\textit{Significance levels: } *** p $<$ 0.001}\\
\end{tabular}
\end{table}

\subsection{Performance on Publication Suitability}
\begin{figure}[!htbp]
    \centering
    \includegraphics[width=\columnwidth]{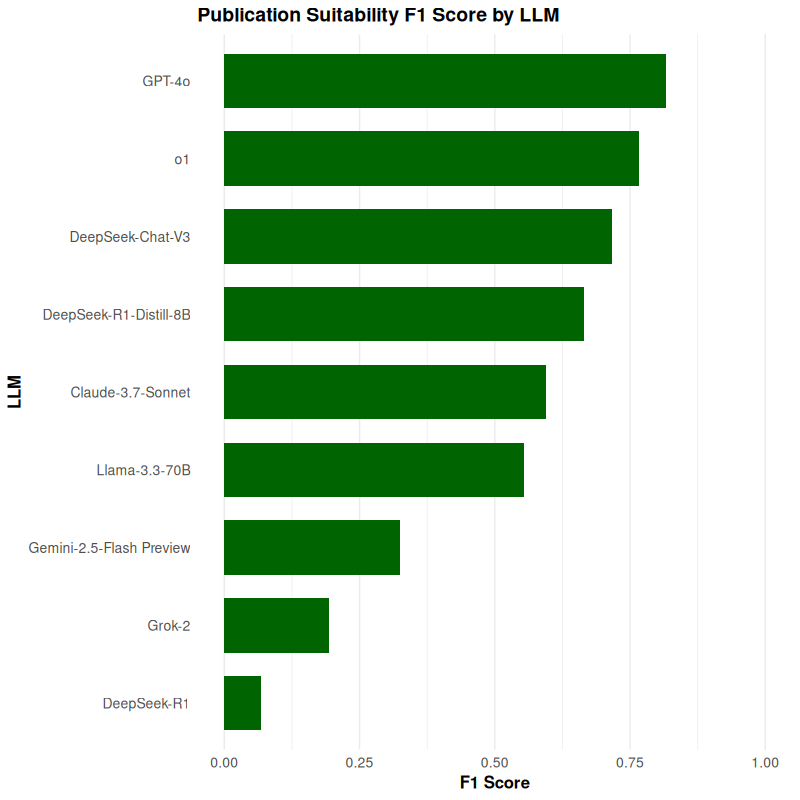}
    \caption{F1 Scores for Publication Suitability by LLM.}
    \label{fig:pub_f1_score}
\end{figure}

Judging overall publication suitability (metrics in Table \ref{tab:pub_suitability_metrics_detailed}) was the most challenging.
Precision scores were generally moderate to high. Recall was a significant issue for several models.
F1 scores (Table \ref{tab:pub_suitability_metrics_detailed} and Figure \ref{fig:pub_f1_score}) varied dramatically, with GPT-4o (0.82) and o1 (0.77) performing best.

\begin{table}[!htbp]
\centering
\caption{Precision, Recall, and F1 Score for Publication Suitability}
\label{tab:pub_suitability_metrics_detailed}
\begin{tabular}[t]{@{}lrrr@{}}
\toprule
\multicolumn{1}{c}{\textbf{LLM}} & \multicolumn{1}{c}{\textbf{Precision}} & \multicolumn{1}{c}{\textbf{Recall}} & \multicolumn{1}{c}{\textbf{F1 Score}}\\
\midrule
Claude-3.7-Sonnet & 0.78 & 0.48 & 0.59\\
DeepSeek-Chat-V3 & 0.78 & 0.66 & 0.72\\
DeepSeek-R1 & 0.91 & 0.03 & 0.07\\
DeepSeek-R1-Distill-8B & 0.50 & 0.99 & 0.66\\
GPT-4o & 0.76 & 0.89 & 0.82\\
Gemini-2.5-Flash Preview & 0.71 & 0.21 & 0.33\\
Grok-2 & 0.76 & 0.11 & 0.19\\
Llama-3.3-70B & 0.79 & 0.43 & 0.55\\
o1 & 0.63 & 0.97 & 0.77\\
\bottomrule
\end{tabular}
\end{table}

The ANOVA for publication suitability (Table \ref{tab:anova_pub_suitability}) shows extremely high F-values and significance (p $<$ 0.001) for all three factors: `llm`, `paper`, and `llm:paper`. The F-value for the `llm` factor (88.539) is particularly large, indicating very substantial differences in how LLMs assess overall suitability. This complex, synthetic judgment task most strongly differentiates the LLMs and is highly dependent on both the LLM's characteristics and the specific paper summary, as well as their interaction. This underscores the difficulty and variability in LLM performance when making holistic evaluations based on synthesized information.

\begin{table}[!htbp]
\centering
\caption{Publication Suitability ANOVA Results}
\label{tab:anova_pub_suitability}
\begin{tabular}[t]{@{}lrrrrl@{}}
\toprule
Source & df & Sum of Squares & Mean Square & F-value & p-value\\
\midrule
llm & 8 & 60.807 & 7.601 & 88.539 & 2.69e-137***\\
model & 57 & 191.397 & 3.358 & 39.114 & 0.00e+00***\\
llm:model & 456 & 571.993 & 1.254 & 14.612 & 0.00e+00***\\
Residuals & 4698 & 403.313 & 0.086 & NA & NA\\
\bottomrule
\multicolumn{6}{l}{\rule{0pt}{1em}\textit{Significance levels: } *** p $<$ 0.001}\\
\end{tabular}
\end{table}

\subsection{Overall Performance}
\begin{figure}[!htbp]
    \centering
    \includegraphics[width=\columnwidth]{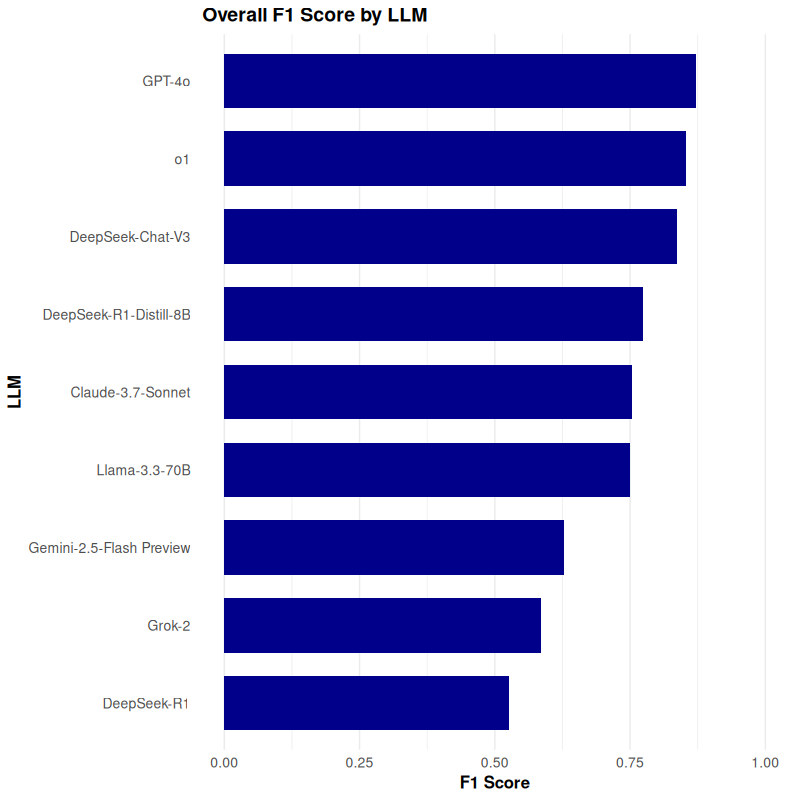}
    \caption{Overall F1 Scores by LLM (averaged across the three criteria).}
    \label{fig:overall_f1_score}
\end{figure}

The overall F1 scores, averaging performance across the three criteria for each LLM, are displayed in Figure \ref{fig:overall_f1_score}. GPT-4o achieved the highest overall F1 score (0.87), followed closely by o1 (0.85) and DeepSeek-Chat-V3 (0.84). DeepSeek-R1 (0.53) and Grok-2 (0.59) had the lowest overall F1 scores.

Further summarizing overall performance, Table \ref{tab:overall_eval_metrics} shows the aggregated Precision, Recall, and F1 Score across all three criteria (calculated from the total true positives, false positives, and false negatives over all decisions). GPT-4o (Overall F1: 0.87), o1 (Overall F1: 0.85), and DeepSeek-Chat-V3 (Overall F1: 0.84) were again the top performers by this aggregate F1 measure. The results highlight a trade-off for some models: DeepSeek-R1-Distill-8B, for instance, had very high overall recall (0.97) but lower overall precision (0.67). Conversely, DeepSeek-R1 had high overall precision (0.92) but very low recall (0.45).

\begin{table}[!htbp]
\centering
\caption{Overall Evaluation Metrics by LLM (Aggregated across criteria)}
\label{tab:overall_eval_metrics}
\begin{tabular}[t]{@{}lrrr@{}}
\toprule
\multicolumn{1}{c}{\textbf{LLM}} & \multicolumn{1}{c}{\textbf{Precision}} & \multicolumn{1}{c}{\textbf{Recall}} & \multicolumn{1}{c}{\textbf{F1 Score}}\\
\midrule
Claude-3.7-Sonnet & 0.85 & 0.68 & 0.75\\
DeepSeek-Chat-V3 & 0.85 & 0.83 & 0.84\\
DeepSeek-R1 & 0.92 & 0.45 & 0.53\\
DeepSeek-R1-Distill-8B & 0.67 & 0.97 & 0.77\\
GPT-4o & 0.84 & 0.91 & 0.87\\
Gemini-2.5-Flash Preview & 0.85 & 0.54 & 0.63\\
Grok-2 & 0.85 & 0.51 & 0.59\\
Llama-3.3-70B & 0.86 & 0.68 & 0.75\\
o1 & 0.77 & 0.98 & 0.85\\
\bottomrule
\end{tabular}
\end{table}

Collectively, these results indicate that while current LLMs are quite capable of assessing the logical coherence of variable choices against a research aim, their performance in evaluating the deeper theoretical plausibility of quantitative results (coefficients) and making nuanced judgments about overall publication suitability varies considerably. Top-performing models like GPT-4o show promise, but significant challenges remain for many others, particularly in tasks requiring more profound domain-specific reasoning. The ANOVA results further emphasize that LLM performance is not uniform but is significantly influenced by the specific LLM, the nature of the paper summary, and the interaction between these two factors, with these effects becoming more pronounced for more complex evaluation criteria.

\section{Discussion}\label{sec:Discussion}
The empirical findings of this study offer a detailed perspective on the capabilities and limitations of current Large Language Models in assessing the economic soundness and theoretical consistency of spatial econometrics research. The observed heterogeneity in performance across different LLMs and evaluation criteria warrants a deeper exploration, potentially linked to their underlying architectures, training, and specific optimizations.

A clear strength across models is the assessment of variable choice coherence. The high F1 scores and the ANOVA results, which show significant but less pronounced F-values for `llm` and `paper` factors compared to other criteria, suggest this task aligns well with the pattern recognition and information synthesis capabilities inherent in most modern LLM architectures, including Dense Transformers (Llama-3.3-70B, DeepSeek-R1-Distill-8B), Mixture-of-Experts (DeepSeek models), Unified (GPT-4o), and Hybrid Reasoning (Claude-3.7-Sonnet). Sufficient context window sizes (128k+) in most evaluated models likely aid in processing the research aim and variable list effectively.

The significant drop in performance for many LLMs when evaluating coefficient plausibility and publication suitability points to the challenges these models face with deeper, domain-specific reasoning. Top-performing models like GPT-4o and o1, both from OpenAI, known for extensive fine-tuning and large (though unspecified for o1 beyond "reasoning-focused") parameter counts, excelled here. GPT-4o's "Unified Architecture" might offer a more holistic integration of information, while o1's explicit "reasoning-focused" design and large 200k context window could directly contribute to its stronger performance in these complex tasks. Similarly, DeepSeek-Chat-V3, a very large MoE model (671B parameters), performed well, possibly benefiting from its sheer scale and capacity to store vast amounts of information, coupled with fine-tuning optimized for chat and instruction-following, which may translate to better adherence to the evaluative prompts.

Conversely, models like DeepSeek-R1, despite their large MoE architecture and "reasoner" designation, performed notably poorer than their chat-tuned sibling, particularly in terms of recall for publication suitability. This suggests that the specific nature of fine-tuning (e.g., chat interaction vs. a more abstract "reasoning" target) can significantly impact performance on evaluative tasks that require interpreting and responding to nuanced instructions. The smaller distilled model, DeepSeek-R1-Distill-8B (8B parameters), showed variable performance; its high recall in some areas but lower precision might reflect the trade-offs of distillation, where general pattern recognition is retained but finer-grained discernment is reduced due to smaller capacity. It must be noted that the higher failure rate of this last model makes it less powerful than the other models, even when the metrics are similar.

Google's Gemini-2.5-Flash Preview, potentially a smaller or faster variant, excelled at variable selection but struggled with the more demanding tasks, aligning with the expectation that scaled-down models might prioritize efficiency over depth of reasoning. Claude-3.7-Sonnet, with its "Hybrid Reasoning Architecture" and 200k context, performed moderately well. Its architecture is designed to dynamically allocate resources, which might explain its competence but perhaps also some variability if the summaries did not always trigger its most intensive reasoning pathways or if its specialized knowledge in spatial econometrics is still developing. Llama-3.3-70B, a 70B parameter Dense Transformer, also showed moderate performance, which is respectable for an open-community model but might be outperformed by proprietary models with potentially larger scale or more targeted fine-tuning for evaluative tasks.

The ANOVA results are crucial in this context. The consistent significance of the `llm` factor across all criteria confirms that inherent model differences (architecture, size, training data, fine-tuning) are primary drivers of performance. The `paper` factor's significance highlights that the characteristics of the research summary itself (complexity, clarity, subtlety of theoretical links) also play a vital role. Most importantly, the highly significant `llm:paper` interaction effect across all tasks, especially pronounced for coefficient plausibility and publication suitability, indicates that no single LLM is universally superior. The "best" LLM often depends on the specific paper summary being evaluated. This suggests that some models might have strengths in processing certain types of information or reasoning about particular theoretical constructs that are present in some paper summaries but not others. For instance, a model with strong pattern-matching might excel with straightforward theoretical checks but falter with more counterintuitive (yet valid) or complex spatial spillover interpretations.

The implications for AI-assisted peer review are thus nuanced. While LLMs can assist with surface-level checks, their reliability for deeper scientific assessment is variable and context-dependent. The strong interaction effects observed suggest that deploying a single LLM for all review tasks might not be optimal. Instead, a more tailored approach, potentially using different LLMs for different sub-tasks or types of papers, might be more effective, though this adds complexity. Ultimately, human oversight remains indispensable, especially given the current limitations in consistent, deep domain reasoning and the opaque nature of LLM decision-making processes. The study's own limitations, such as using summaries and the specific design of counterfactuals, mean these observations are a snapshot, but they underscore the need for ongoing, critical evaluation of LLM capabilities as they evolve.

\section{Conclusion and Future Directions}\label{sec:Conclusion}
This paper has systematically investigated the capabilities of a diverse set of Large Language Models in assessing key aspects of economic soundness and theoretical consistency within the specialized field of spatial econometrics research. By employing a methodology based on original and counterfactual paper summaries and incorporating ANOVA to analyze performance variations, we found that LLMs demonstrate proficiency in evaluating the surface-level coherence of variable choices. However, they face significant challenges in tasks requiring deeper economic reasoning, such as assessing the plausibility of coefficients and judging overall publication suitability, with performance varying considerably among models.

The ANOVA results consistently highlighted that LLM performance is significantly influenced not only by the choice of the LLM itself—reflecting differences in architecture, scale, and training—and the intrinsic characteristics of the paper summary being evaluated, but also, crucially, by the interaction between these two factors. This indicates that no single LLM excels universally; their effectiveness is context-dependent, varying with the specific research content they are tasked to assess. This complex interplay, especially for more nuanced evaluation criteria, underscores the current limitations of LLMs in autonomously conducting comprehensive peer review in quantitative disciplines. While they show promise as assistive tools, particularly for initial screening, their outputs for more complex judgments necessitate careful human oversight and validation. The journey towards reliable AI-assisted peer review will require further advancements in LLM reasoning capabilities, particularly in specialized domains, and a deeper understanding of how to best integrate them into human-centric evaluation workflows, taking into account these significant interaction effects.

Future research should continue to explore these dynamics. Investigating LLM performance with full-text access, rather than summaries, could reveal different capabilities and limitations, allowing models to leverage broader contextual information. Expanding the evaluation criteria to include the methodological rigor of spatial econometric techniques, the novelty of contributions, and the appropriateness of data would provide a more holistic picture of their potential role in peer review. Continuous benchmarking of new and evolving LLM architectures, including those specifically fine-tuned for scientific or economic text, remains essential as the technology rapidly advances. Further research could also focus on developing interactive review systems where LLMs act as sophisticated assistants to human reviewers, for example, by verifying specific claims, cross-referencing literature, or flagging potential areas of concern for human inspection. A deeper qualitative analysis of LLM reasoning patterns, particularly in instances highlighted by the interaction effects, could offer valuable insights into why certain LLMs struggle with specific types of content or theoretical constructs. Finally, understanding and mitigating potential biases in LLM assessments, and further exploring the nature and implications of the `llm:paper` interaction to potentially guide the selection of appropriate models for specific tasks or paper types, are crucial steps for the responsible development and deployment of these technologies in academic settings.

\begin{flushleft}
The input data and source code used for the paper is available at:
\par
https://github.com/lmorandini/impact\_econometrics
\end{flushleft}

\clearpage
\appendix 
\section*{Appendix: LLM Prompts} \label{app:prompts} 
\subsection*{A) Prompt for Phase 1 (Qualitative Economic Assessment)}
\begin{verbatim}
You are an expert in applied econometrics and spatial economics.
You are given the results of a scientific paper.

Based on the provided description of the aim, target variable, explanatory variables,
and the estimated coefficients, perform a qualitative economic assessment of the model.

In particular:
- Provide an interpretation of the most relevant parameters.
- Evaluate whether the model specification is economically meaningful.
- Assess whether the signs and statistical significance of the coefficients are
  consistent with economic theory.
- Highlight any inconsistencies, surprises, or aspects that might require further
  justification from the authors.
\end{verbatim}

\subsection*{B) Prompt for Phase 2 (Structured JSON Output)}
\begin{verbatim}
Based on this evaluation of a spatial econometric model, produce a structured JSON
object summarizing your evaluation. The JSON should have the following structure:

{
  "economic_sense_of_variable_choice": 1 or 0,
  "economic_sense_of_significant_coefficients": 1 or 0,
  "suitable_for_publication_in_journal": 1 or 0
}

For each element:
- Set the value to 1 (true) if the condition holds, 0 (false) otherwise.
- Base your judgment strictly on the economic rationale of the model, the
  statistical significance and signs of the estimated coefficients, and the
  standards of the Journal [implied based on the context of the papers selected].
Do not explain the JSON. Just return the JSON object.
\end{verbatim}
\clearpage


\bibliographystyle{chicago} 
\bibliography{bibliography}

\vfill 

\end{document}